\documentclass[doublecol]{epl2}

\usepackage{amssymb,amsmath,latexsym}

\newcommand\beq{\begin{equation}}
\newcommand\eeq{\end{equation}}

\newcommand\bea{\begin{eqnarray}}
\newcommand\eea{\end{eqnarray}}

\newcommand\bi{\begin{itemize}}
\newcommand\ei{\end{itemize}}

\newcommand\non{\nonumber}

\newcommand\up{\uparrow}
\newcommand\dn{\downarrow}

\newcommand\epp{e^{i\,(k\,+\,k_F)\,x}}
\newcommand\epm{e^{i\,(-k\,+\,k_F)\,x}}

\newcommand\emp{e^{-i\,(k\,+\,k_F)\,x}}
\newcommand\emm{e^{-i\,(-k\,+\,k_F)\,x}}

\newcommand\hmi{{\mathcal H}_{\textsf{int}}}

\newcommand\odd{1{\textendash}D}

\newcommand\od{1{\textendash}D~}

\newcommand\td{2{\textendash}D~}

\newcommand\ie{{\it{i.e.}}}

\newcommand\rgd{{\textsf{RG}}}
\newcommand\qwd{{\textsf{QW}}}

\newcommand\nsnd{{\textsf{NSN}}}
\newcommand\fsnd{{\textsf{FSN}}}
\newcommand\fsfd{{\textsf{FSF}}}
\newcommand\fnfd{{\textsf{FNF}}}

\newcommand\hfd{{\textsf{HF}}}

\newcommand\scurrentd{{\textsf{SC}}}

\newcommand\card{{\textsf{CAR}}}
\newcommand\ard{{\textsf{AR}}}
\newcommand\ndcd{{\textsf{NDC}}}

\newcommand\ed{{\textsf{e}}}
\newcommand\hd{{\textsf{h}}}

\newcommand\wirg{{\textsf{WIRG~}}}
\newcommand\ns{{\textsf{NS~}}}
\newcommand\rg{{\textsf{RG~}}}
\newcommand\qw{{\textsf{QW~}}}

\newcommand\nsn{{\textsf{NSN~}}}
\newcommand\fsn{{\textsf{FSN~}}}
\newcommand\fsf{{\textsf{FSF~}}}

\newcommand\hf{{\textsf{HF~}}}

\newcommand\scurrent{{\textsf{SC~}}} 
\newcommand\tmr{{\textsf{TMR~}}}
\newcommand\car{{\textsf{CAR~}}}
\newcommand\ar{{\textsf{AR~}}}
\newcommand\ndc{{\textsf{NDC~}}}

\newcommand\e{{\textsf{e~}}}

\def\dfrac#1#2{{\displaystyle\frac{#1}{#2}}}

\newif\ifboo \boofalse
\def\Review#1{\boofalse{\it #1},}

\def\Vol#1{\ifboo Vol. {\bf #1}\else{\bf #1}\fi}
\def\Year#1{\ifboo #1\else(#1)\fi}

\def\Page#1{\ifboo {\rm p. #1}\else{\rm #1}\fi}

\title{Spintronics with \nsn Junction of one-dimensional quantum wires\,:
A study of Pure Spin Current and Magnetoresistance} \shorttitle{\nsn
Junction of one-dimensional quantum wires}

\author{Sourin Das\inst{1} \and Sumathi Rao\inst{2} \and Arijit Saha\inst{2}}
\shortauthor{Sourin Das \etal}

\institute{
  \inst{1} {Department of Condensed Matter Physics, Weizmann
Institute of Science, Rehovot 76 100, Israel}\\
  \inst{2} {Harish$-$Chandra  Research Institute,
 Chhatnag Road, Jhusi, Allahabad 211 019, India}
}

\abstract{We demonstrate possible scenarios for production of pure
spin current and large tunnelling magnetoresistance ratios from
elastic co-tunnelling and crossed Andreev reflection across a
superconducting junction comprising of normal
metal-superconductor-normal metal, where, the normal metal is a
one-dimensional interacting quantum wire. We show that there are
fixed points in the theory which correspond to the case of pure
spin current. We analyze the influence of electron-electron
interaction and see how it stabilizes or de-stabilizes the
production of pure spin current. These fixed points can be of
direct experimental relevance for spintronics application of
normal metal-superconductor-normal metal junctions of
one-dimensional quantum wires. We also calculate the power law
temperature dependence of the {\textit crossed Andreev reflection
enhanced} tunnelling magnetoresistance ratio for the normal
metal-superconductor-normal metal junction. }

\begin{document}

\maketitle

\section{Introduction} Two fundamental degrees of freedom associated
with an electron that are of direct interest to condensed matter
physics are its charge and spin. Until very recently, all
conventional electron-based devices have been solely based upon
the utilization and manipulation of the charge degree of freedom
of an electron. However, the realization of the fact that devices
based on the spin degree of freedom can be almost dissipation-less
and with very fast switching times, has led to an upsurge in
research activity in this direction in recent
years~\cite{dattadas,review,rashba}. The first step towards
realization of spin-based electronics (spintronics) would be to
produce pure spin current (\scurrentd). From a purely theoretical
point of view, it is straight forward to define a charge current
as a product of local charge density with the charge velocity, but
such a definition cannot be straight forwardly extended to the
case of \scurrentd. This is because both spin $\vec S$ and
velocity $\vec v$ are vector quantities and hence the product of
two such vectors will be a tensor.

In this letter, we adopt the simple minded definition of
\scurrentd, which is commonly used~\cite{marcus}. It is just the
product of the local spin polarization density associated with the
electron or hole, (a scalar $s$ which is either positive for
up-spin or negative for down-spin)  and its
velocity~\cite{marcus}. The most obvious scenario in which one can
generate a pure \scurrent in the sense defined above would be to
have {\it{(a)}} an equal and opposite flow of identically
spin-polarized electrons through a channel, such that the net
charge current through the channel is nullified leaving behind a
pure \scurrentd, or {\it{(b)}} alternatively, an equal flow of
identically spin polarized electrons and holes in the same
direction through a channel giving rise to pure \scurrent with
perfect cancellation of charge current. In this letter, we explore
the second possibility for generating pure \scurrent using a
normal metal$-$superconductor$-$normal metal (\nsnd) junction.
%
\section{Proposed device and its theoretical modelling}
\begin{figure*}
\begin{center}
\includegraphics[width=12cm,height=5cm]{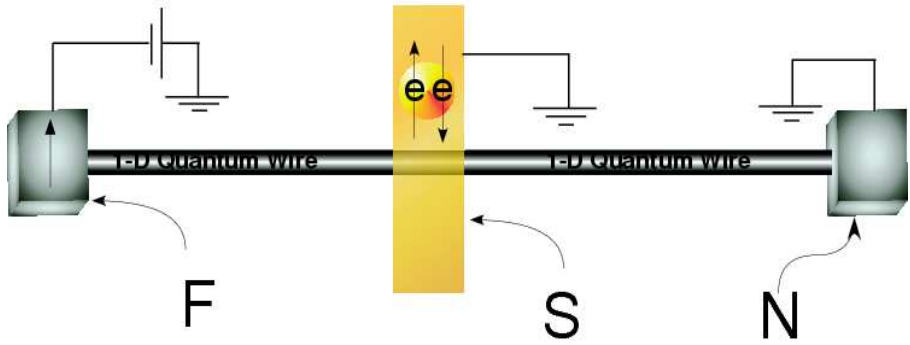}
\caption{\footnotesize{A \od quantum wire (carbon nanotube)
connected to a ferromagnetic (F) lead on the left and a normal (N)
lead on the right. The thin strip in the middle of the wire
depicts a \td layer of superconducting material deposited on top
of the wire.}} \label{geom1}
\end{center}
\end{figure*}

The configuration we have in mind for the production of pure
\scurrent is shown in Fig.~\ref{geom1}. The idea is to induce a
pair potential in a small region of a quantum wire (\qwd) by
depositing a superconducting strip on top of the wire (which may
be, for instance, a carbon nanotube) due to proximity effects. If
the strip width on the wire is of the order of the phase coherence
length of the superconductor, then both direct electron to
electron co-tunnelling as well as crossed Andreev electron to hole
tunnelling can occur across the superconducting
region~\cite{hekking2}. It is worth pointing out that in the case
of a singlet superconductor, which is the case we consider, both
the tunnelling processes will conserve spin.  In order to describe
the mode of operation of the device (see Fig.~\ref{geom1}), we
first assume that the $S$-matrix representing the \nsn junction
described above respects parity symmetry about the junction,
particle-hole symmetry and spin-rotation symmetry. Considering all
the symmetries, we can describe the superconducting junction
connecting the two half wires by an $S$-matrix with only four
independent parameters namely, {\it{(i)}} the normal refection
 amplitude ($r$) for \e (\hd), {\it{(ii)}} the transmission
amplitude ($t$) for \e (\hd), {\it{(iii)}} the Andreev reflection
(\ard) amplitude ($r_A$) for \e (\hd), and {\it{(iv)}} the crossed
Andreev reflection (\card) amplitude ($t_A$) for \e (\hd). If we
inject spin polarized electron ($\up$ \ed)  from the left \qw
using a ferromagnetic contact and tune the junction parameters
such that $t$ and $t_a$ are equal to each other, it will lead to a
pure \scurrent flowing in the right \qw (see Fig.~\ref{geom1}).
This is so because, on an  average, an equal number of electrons
($\up$ \ed) (direct electron to electron tunnelling) and holes
($\up$ \hd) (crossed Andreev electron to hole tunnelling) are
injected from the left wire to the right wire resulting in
production of pure \scurrent in the right wire. Note that spin up
holes ($\up$ \hd) implies a Fermi sea with an absence of spin down
electron (which is what is needed for the incident electron $(\up
\ed)$ to form a Cooper pair and enter the singlet superconductor).
\section{Renormalisation Group study}
 We now include the effects of inter-electron interactions on the
$S$-matrix using the renormalisation group (\rgd) method
introduced in Ref.~\cite{yue}. This was furthur generalized to the
case of multiple wires in Refs.~\cite{lal,das} and to the case of
\od \ns junction~\cite{jap3,man,titov}. The basic idea of the
method is as follows. The presence of scattering (reflection)
induces Friedel oscillations in the density of non-interacting
electrons. Within a mean field picture for a weakly interacting
electron gas, the electron not only scatters off the potential
barrier but also scatters off these  density oscillations with an
amplitude proportional to the interaction strength. Hence by
calculating the total reflection amplitude due to scattering from
the scalar potential scatterer and from the Friedel oscillations
created by the scatterer, we can include the effect of
electron-electron interaction in calculating transport. This
approach can be generalized to junctions of one-dimensional (\odd)
\qw with a single superconductor. In this case, there will be
non-zero \ar in the bulk of the wire due to proximity induced pair
potential, besides the \ar right at the \nsn junction which turns
an incoming electron into an outgoing hole.

 {The \rg equations for a \ns case have been obtained earlier using
bosonization \cite {jap1,fazio,winkelholz} and using \wirg
\cite{jap3,titov}. In this letter we extend these \wirg results to
the \nsn case.} To obtain the \rg equations in the presence of \ar
and \car for the \nsn junction, we follow a procedure similar to
that used in Ref.~\cite{lal}. The fermion fields expanded around the
left and right Fermi momenta on each wire can be written as, $
\psi_{is} (x) = \Psi_{I\,is}(x)\, e^{i\,k_F\,x} \,+\,
\Psi_{O\,is}(x)\, e^{-i\,k_F\,x} ~; $
where $i$ is the wire index, $s$ is the spin index which can be
$\up,\dn$ and $I(O)$ stand for incoming (outgoing) fields. Note
that $\Psi_{I(O)}(x) $ are slowly varying fields on the scale of
$k_F^{-1}$. For a momentum in the vicinity of $k_F$, the incoming
and outgoing fields can be Fourier expanded as:
\bea
\Psi_{ks}(x) &=& \int \,dk\, \Big[\, b_{ks} \,\epp \,+\,
d_{ks}^\dagger \, \epm  \non\\
 &&\,+\, r \,b_{ks}\, \emp  \,+
 \, r^\star \,d_{ks}^\dagger \, \emm  \non\\
 && \,+\, r_A \,d_{ks}\, \emm  \,+\, r_A^\star \,b_{ks}^\dagger\, \emp
\Big]
 \non\\
 \label{equation1}
\eea
%
where $b_{ks}^{}$ is the electron destruction operator and
$d_{ks}^{}$ is the hole destruction operator and we have allowed for non-
conservation of charge due to the proximity effect.  We allow for
short-range density-density interactions between the fermions,
$ \hmi = \dfrac{1}{2} \, \int dx \,dy \,\rho_{is}^{} V(x-y)
\,\rho_{is}^{} $,
\noindent where the sum over the spin indices is assumed.

Then the effective Hamiltonian, can be derived using a
Hartree$-$Fock (\hfd) decomposition of the interaction. The charge
conserving \hf decomposition leads to the interaction Hamiltonian
(normal) of the form
\bea
\hmi^N &=&
\frac{-i(g_2-2g_1)}{4\pi}
\int_0^\infty
\frac{dx}{x}
\Bigg[r_i^\star \big(\Psi_{I\,i\up}^{\dagger} \Psi_{O\,i\up}^{} +
\non\\
&&
\Psi_{I\,i\dn}^{\dagger}  \Psi_{O\,i\dn}^{} \big)
 -
 r_i \left( \Psi_{O\,i\up}^{\dagger}
\Psi_{I\,i\up}^{}
+ \Psi_{O\,i\dn}^{\dagger}  \Psi_{I\,i\dn}^{}
\right)
\Bigg]
\non\\
\eea
(We have assumed spin-symmetry and used $r_{i\up} = r_{i\dn} = r_i$.)
\begin{figure}
\begin{center}
\includegraphics[width=8.5cm,height=7.5cm]{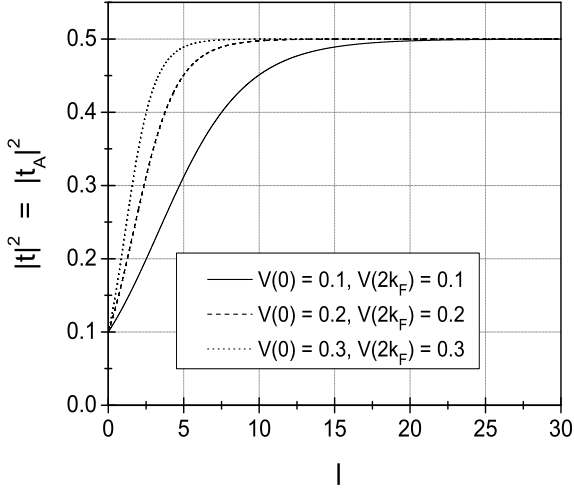}
\caption{\footnotesize{The variation of $|t|^2$ (=$|t_A|^2$) is
plotted as a function of the  {dimensionless parameter $l$ where
$l=ln(L/d)$ and $L$ is either $L_{T}=\hbar v_{F}/k_{B} T$ at zero
bias or $L_{V}=\hbar v_{F}/eV$ at zero temperature and $d$ is the
short distance cut-off for the \rg flow.} The three curves
correspond to three different values of $V(0)$ and $V(2k_{F})$ for
the \nsn junction. This case corresponds to the $S$-matrix given by
$S_1$.}}
 \label{nsnfig2}
\end{center}
\end{figure}
This has been derived earlier in Ref.~\cite{lal}. We use the same
method, but now, we also allow for a charge non-conserving \hf
decomposition and arrive at the (Andreev) Hamiltonian
\bea \hmi^A
&=&
\frac{-i(g_1+g_2)}{4\pi}
\int_0^\infty
\frac{dx}{x}
 \Bigg[-\,r_{Ai}^\star  \big(
 \Psi_{I\,i\up}^{\dagger} \Psi_{O\,i\dn}^\dagger
+ \non\\
&&
 \Psi_{O\,i\up}^{\dagger} \Psi_{I\,i\dn}^\dagger \big)
+ r_{Ai}  \left(\Psi_{O\,i\dn}^{} \Psi_{I\,i\up}^{}
+
\Psi_{I\,i\dn}^{}  \Psi_{O\,i\up}^{}\right) \Bigg]
\non\\
 \eea
Note that, even if this appears non-charge conserving, charge
conservation is taken care of by the $2e$ charge that flows into the
superconductor due to proximity induced Cooper pairing.

In Ref.~\cite{lal}, the perturbatively calculated correction to
the reflection amplitude under $\exp{[-i \hmi^N \,t]}$ was derived
to first order in $\alpha$. For electrons with spin incident with
momentum $k$ with respect to $k_F$, this was shown to be given by
$ \dfrac{-\alpha \, r_{i}}{2} \, \ln (kd) \label{Friedeln} $
where $\alpha =(g_2-2g_1) / 2\pi\hbar v_F$, $d$ is a short
distance cutoff and $g_1=V(2k_F)$, $g_2=V(0)$. It is worth noting
that both $g_1$ and  $g_2$ scale under \rgd~\cite{yue}. Hence, the
values given above for $g_1$ and $g_2$ are the microscopic values
it takes in the original Hamiltonian defined at the short distance
scale.

Analogously, here, we calculate the amplitude to go from an
incoming electron wave to an outgoing hole wave under $\exp{[-i
\hmi^A \, t]}$. It is given by
$\dfrac{\alpha^\prime \, r_{A\,i}}{2} \, {\ln (kd)}
\label{FriedelA} $
 where $\alpha^\prime = (g_1+g_2)/ 2 \pi \hbar v_F$.
Note that the spin of the outgoing hole is always the same as the
spin of the incoming electron, since the Andreev Hamiltonian also
conserves spin for a singlet superconductor. We see that there is
a logarithmic singularity at the $k \to 0$ limit which implies
that the lowest order perturbation theory is not enough to
calculate correction to the reflection and \ar amplitudes when the
momenta of the incident particles are very close to the Fermi wave
vector. Following Yue \etal~\cite{yue}, we sum up these most
divergent processes using the "poor man's scaling"
approach~\cite{anderson} to obtain \rg equations for the normal
refection amplitude ($r$), the transmission amplitude ($t$), the
\ar amplitude ($r_A$), and the \car amplitude ($t_A$)  which are
as follows,

\bea
\frac{dr}{dl} &=& -\,\Big[\frac{\alpha}{2} \,\left\{\left(t^2
\,+\, r_{A}^2 \,+\, t_{A}^2\right) \,r^\star \,-\, r
\,\left(1-|r|^2\right)\right\} \non\\
&& \,-\, \alpha^\prime\,\left\{r\, |r_{A}|^2 \,+\, r_{A}^\star
\,t_{A} \,t\right\}\Big] \label{rnsn}
\\
%
%
\frac{dr_{A}}{dl} &=&
-\,\Big[\alpha\,\left\{|r|^2r_{A}\,+\,t\,t_{A} \,r^\star\right\}
\,+\, \frac{\alpha^\prime}{2}\,\big\{r_{A}
 \,
\non\\
&&
 -\,\left(r^2
\,+\,r_{A}^2\,+\,t^2\,+\,t_{A}^2\right)\,r_{A}^\star \big\}\Big]
\label{ransn}
%
\\
\frac {dt}{dl} &=&-\,\Big[\alpha \,\left\{|r|^2\,t \,+\, r^\star
\,r_A\,t_A\right\} \non\\ &&
 \,-\, \alpha^\prime \, \left\{ |r_A|^2
t \,+\, r\, r_A^\star \, t_A\right\}\Big] \label{tnsn}
\\
\frac{dt_A}{dl} &=& -\,\Big[\alpha \, \left\{r^\star \, r_A\,
t\,+\,|r|^2\, t_A\right\} \non\\ &&
 \,-\, \alpha^{\prime}\, \left\{r\,t \,r_A^\star
\,+\,|r_A|^2\,t_A\right\}\Big] \label{tansn}
\eea  {Note that when $t=t_{A}=0$ these equations reduce to the \rg
equations obtained in Ref. \cite {jap3} for the case of \ns
junction.}

\begin{figure*}
\includegraphics[width=8.9cm,height=8.5cm]{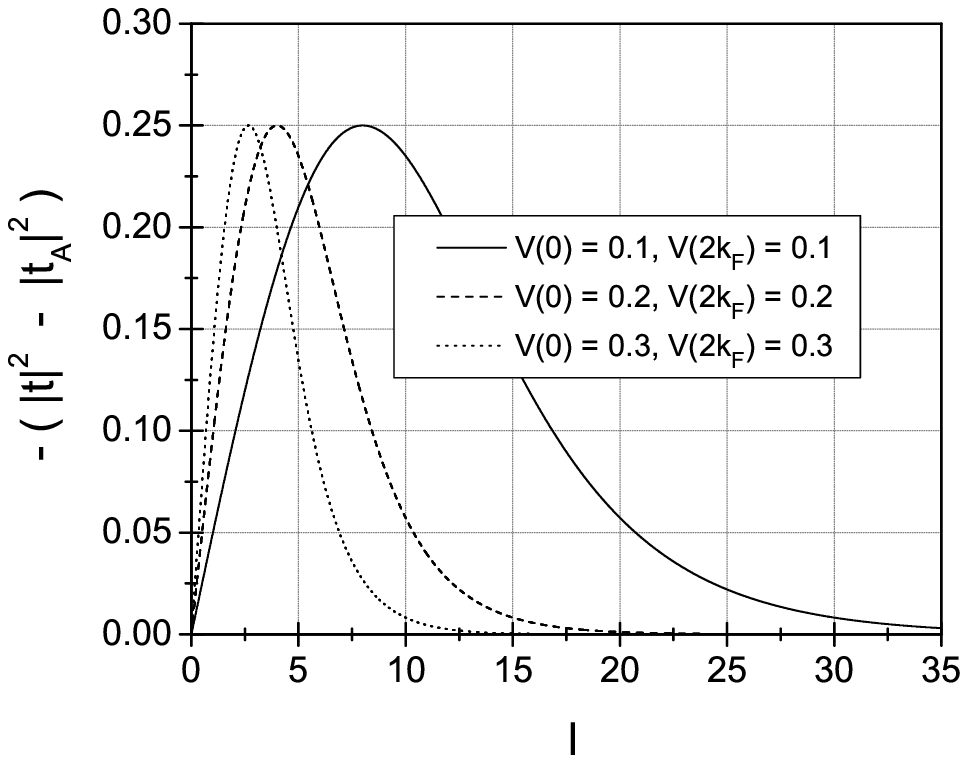}
\vskip
-8.5cm \hskip 3.4in
\includegraphics[width=8.9cm,height=8.5cm]{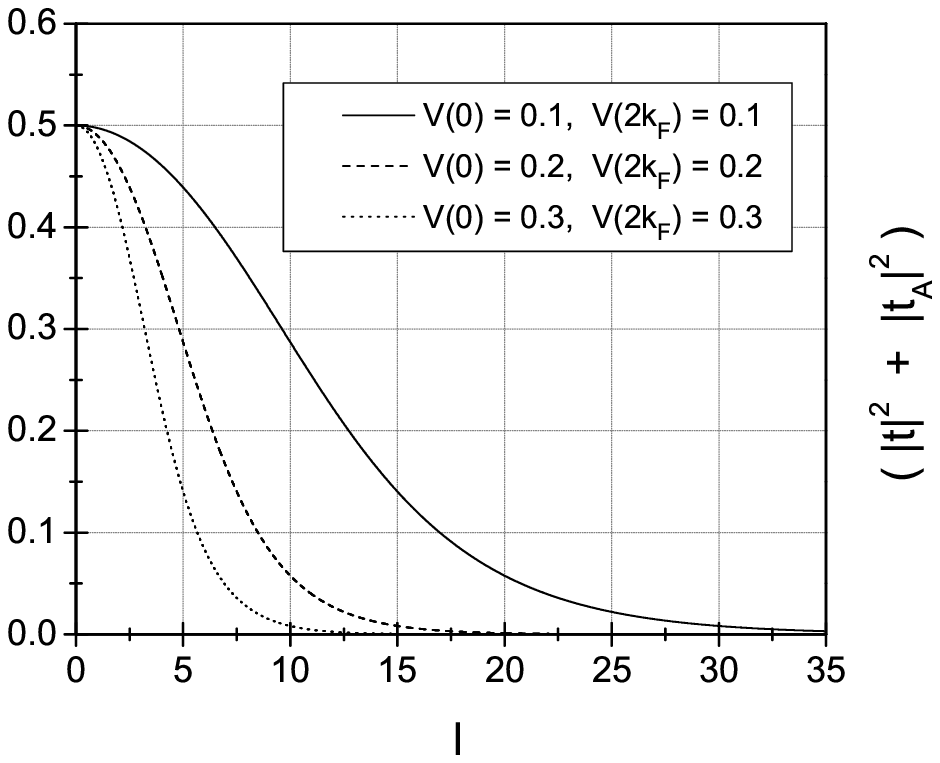}
\caption{\footnotesize{The variation of $-(|t|^2 - |t_A|^2)$
$\propto G^{C}_{\up}$ or $G^{C}_{\dn}$ and the variation of $(|t|^2
+ |t_A|^2)$ $\propto G^{S}_{\up}$ or $G^{S}_{\dn}$ are plotted as a
function of the  {dimensionless parameter $l$ where $l=ln(L/d)$ and
$L$ is either $L_{T}=\hbar v_{F}/k_{B} T$ at zero bias or
$L_{V}=\hbar v_{F}/eV$ at zero temperature and $d$ is the short
distance cut-off for the \rg flow.} The three curves in each plot
correspond to three different values of $V(0)$ and $V(2k_{F})$ for
the \nsn junction. These plots correspond to the $S$-matrix given by
$S_2$. }}
\label{nsnfig3}
\end{figure*}
\section{Results and Discussion}
We propose two possible $S$-matrices ($S_1$ and $S_2$) that can be
realized within our set-up which  will lead to production of pure
\scurrentd. The {\it{spin conductance}} is defined as $G^{S}_{\up}
(G^{S}_{\dn}) \propto |t|^2 + |t_A|^2$ whereas the {\it{charge
conductance}} is given by $G^{C}_{\up} (G^{C}_{\dn}) \propto
-(|t|^2-|t_A|^2)$. The $\up$ and $\dn$ arrows in the subscript
represent the spin polarization of the injected electrons from the
ferromagnetic lead (see Fig.~\ref{geom1}). The negative sign in
the expression for $G^{C}_{\up} (G^{C}_{\dn})$ arises because it
is a sum of contribution coming from two oppositely charged
particles (electrons and holes). The first $S$-matrix, $S_1$ has
$r=0$ (reflection-less), $r_A \neq 0$ and $t=t_A$. This is not a
fixed point and hence the parameters of the $S$-matrix will flow
under \rg. It is easy to see from Eqs.~\ref{rnsn} - \ref{tansn},
that for this case, the \rg equations for $t$ and $t_A$ are
identical, and hence it is ensured that the \rg flow will retain
the equality of the $t$ and $t_A$ leading to the preservation of
pure \scurrent. Physically this implies that  if we start the
experiment with this given $S$-matrix ($S_1$) at the high energy
scale (at finite bias voltage and zero temperature {\it{or}} at
zero bias and finite temperature), then, as we reduce the bias in
the zero temperature case (or reduce the temperature in the zero
bias case), the correlations arising due to inter-electron
interactions in the wire are such that the amplitude of $t$ and
$t_A$ will remain equal to each other. The quantity which
increases with increasing length scale $L$ is the absolute value
of the amplitude $t$ or $t_A$ leading to a monotonic increase of
pure \scurrent till it saturates at the maximum value allowed by
the symmetries of the $S$-matrix, $S_1$ (Fig.~\ref{nsnfig2}). Here
all the $S$-matrix elements are assumed to be energy independent
and hence the bias dependence is solely due to \rg flow. Of course
the bias window has to be small enough so that the energy
dependence of $t$, $t_A$, $r$ and $r_A$ can be safely ignored.
This saturation point is actually a stable fixed point of the
theory if the junction remains reflection-less. So we observe that
the transmission (both $t$ and $t_A$) increases to maximum value
while the \ar amplitude scales down to zero.
This flow direction is quite different from that of the standard
case of a single impurity in an interacting electron gas in \od
where any small but finite reflection amplitude gets enhanced
under \rg flow ultimately leading to zero transmission~\cite{k&f}.
The difference here is because the  \rg flow is solely due to the
existence of the finite pair potential (due to $r_A$) and not due
to the usual Friedel oscillations (due to $r$).
Hence the electrons in the wire have an effective attractive
interaction leading to a counter intuitive \rg flow.
We remark that the interaction induced correction enhances the
amplitude for pure \scurrent and also stabilizes the pure
\scurrent operating point. This makes the operating point, $S_1$
quite well-suited for an experimental situation.
Fig.~\ref{nsnfig2} shows the variation of the pure spin
conductance ($= 2 \times |t|^2$ in units of $e^2/h$) as a function
of relevant length scale, $L$ of the problem.

The second case corresponds to the most symmetric $S$-matrix
($S_2$). It is a fixed point of \rg equations and is given by
$r_{}=1/2, r_{A}=-1/2, t_{}=1/2, t_{A}=1/2$.
Here also $t$ is equal to
$t_A$ as in the previous case and thus the junction will act like
a perfect charge filter resulting in pure \scurrent in the right
wire (if spin polarized charge current is injected in the left
wire). However, this $S$-matrix ($S_2$) represents an unstable
fixed point. Due to any small perturbation, the parameters tend to
flow away from this unstable fixed point to the most stable
disconnected fixed point given by $|r|=1$ as a result of \rg flow.
So this $S$-matrix ($S_2$) is not a stable operating point for the
production of pure \scurrentd. But, it is interesting to note that
if we switch on a small perturbation around this fixed point, the
charge conductance exhibits a non-monotonic behavior under \rg
flow (Fig.~\ref{nsnfig3}). This non-monotonicity results from two
competing effects {\textit {viz.}}, transport through both
electron and hole channels and, the \rg flow of $g_1$, $g_2$. This
essentially leads to negative differential conductance
(\ndcd)~\cite{esaki}.
 Elaborating it further, all it means is that if we start an
 experiment with this given $S$-matrix ($S_2$) at zero
temperature and at finite bias, then as we go towards zero
bias,the conductance will show a rise with decreasing bias for a
certain bias window. This can be seen from Fig.~\ref{nsnfig3}.
This aspect of the \rg flow can be of direct relevance for
manipulating electron and spin transport in some
mesoscopic devices.
\begin{figure*}
\includegraphics[width=8.9cm,height=8.5cm]{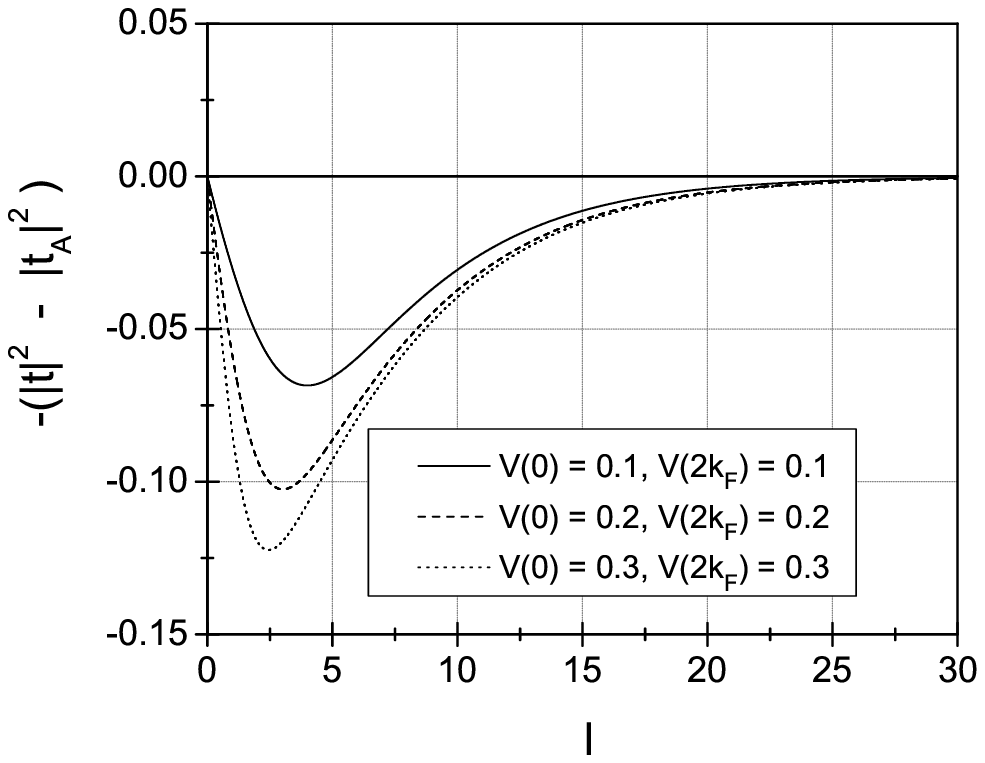}
 \vskip
-8.5cm \hskip 3.4in
\includegraphics[width=8.9cm,height=8.5cm]{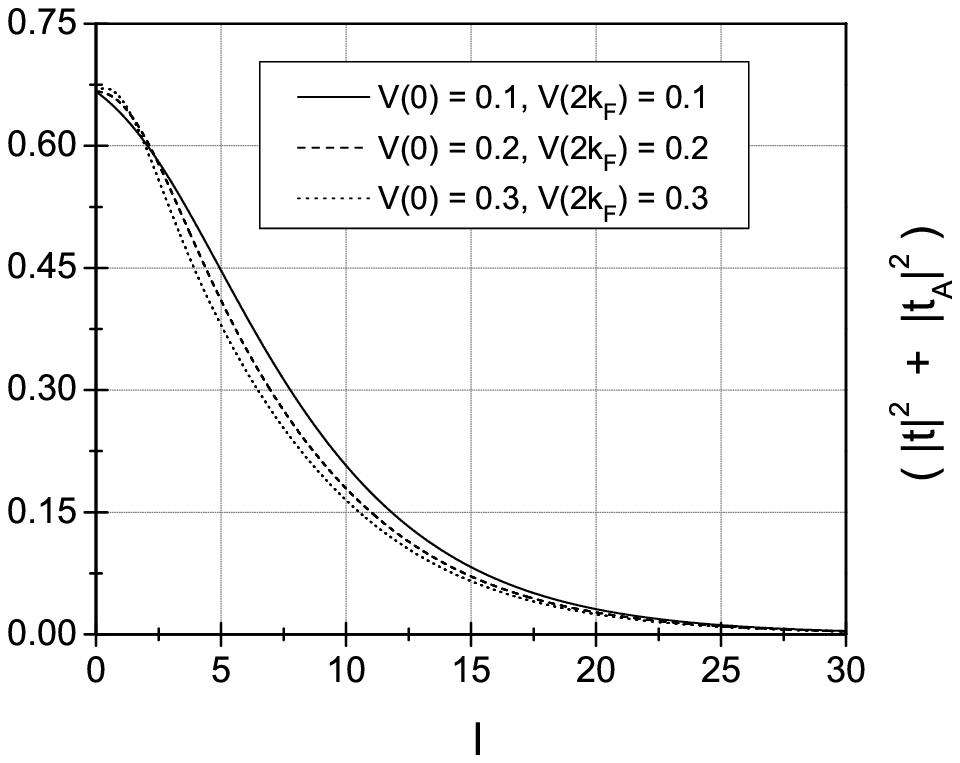}
\caption{\footnotesize{The variation of $-(|t|^2 - |t_A|^2)$
$\propto G^{C}_{\up}$ or $G^{C}_{\dn}$ and the variation of $(|t|^2
+ |t_A|^2)$ $\propto G^{S}_{\up}$ or $G^{S}_{\dn}$ are plotted in
left and right panel plots as a function of the {dimensionless
parameter $l$ where $l=ln(L/d)$ and $L$ is either $L_{T}=\hbar
v_{F}/k_{B} T$ at zero bias or $L_{V}=\hbar v_{F}/eV$ at zero
temperature and $d$ is the short distance cut-off for the \rg flow.}
The three curves in each plot correspond to three different values
of $V(0)$ and $V(2k_{F})$ for the \fsn junction. These plots
correspond to the $S$-matrix given by $S_3$.}}
\label{nsnfig4}
\end{figure*}

Now we will switch to the case of ferromagnetic half
metal$-$superconductor$-$normal metal (\fsnd) junction which
comprises of a \od ferromagnetic half metal (assuming $\up$
polarization) on one side and a normal \od metal on the other side
(in a way similar to the set-up shown in Fig.~\ref{geom1}). This
case is very complicated to study theoretically because  the
minimal number of independent complex-valued parameters that are
required to parameterize the $S$-matrix is nine as opposed to the
previous (symmetric) case which had only four such parameters.
These are given by $r_{\up\up}^{11}, ~r_{\up\up}^{22},
~r_{\dn\dn}^{22}, ~t_{A\up\up}^{12}, ~t_{A\dn\dn}^{21},
~r_{A\up\up}^{22}, ~r_{A\dn\dn}^{22}, ~t_{\up\up}^{12},
~{\mathrm{and}}~ t_{\up\up}^{21}$.
 Here, $1$($2$) is the wire index for the ferromagnetic
 (normal) wire while, $\up$ and $\dn$ are the respective
 spin polarization indices for the electron.
%
The large number of independent parameters in this case arise
because of the presence of ferromagnetic half-metallic wire which
destroys both the spin rotation symmetry and the left-right
symmetry. The only remaining symmetry is the particle-hole
symmetry. Analogous to the \rg equations (given by
Eqs.~\ref{rnsn}-\ref{tansn}) for the \nsn case, it is possible to
write down all the nine \rg equations for \fsn case and solve them
numerically to obtain the results as shown in Fig.~\ref{nsnfig4}.
(For further details, see Ref.~\cite{drsaha}). In this case, the
elements of a representative $S$-matrix ($S_3$) which correspond
to the production of pure \scurrent are $ |r_{\up\up}^{11}|=
|r_{\up\up}^{22}|=|r_{\dn\dn}^{22}|=|t_{A\up\up}^{12}|=
|t_{A\dn\dn}^{21}|=|r_{A\up\up}^{22}|=|r_{A\dn\dn}^{22}|
=|t_{\up\up}^{12}|=\lvert t_{\up\up}^{21}\rvert=1/\sqrt{3}$
and the corresponding phases associated with each of these
amplitudes are $\pi/3,\pi,0,-\pi/3,0,\pi/3,0,\pi,-\pi/3$
respectively. By solving the nine coupled \rg equations for the above
mentioned nine independent parameters, we have checked numerically that
this is {\it{not}} a fixed point of the \rg equation and hence it will
flow under \rg and finally reach the trivial stable fixed point
given by $r_{\up\up}^{11} = r_{\up\up}^{22} = r_{\dn\dn}^{22} =
1$.
Now if we impose a bias on the system from left to right, it  will
create a pure \scurrent on the right wire because
$|t_{A\up\up}^{12}|$ is exactly  equal to $|t_{\up\up}^{12}|$. But
of course, this is a highly unstable operating point for
production of pure \scurrent as this is not even a fixed point and
hence will always flow under any variation of temperature or bias
destroying the production of pure \scurrentd. In this case also,
the spin conductance shows a monotonic behavior while, the charge
conductance is non-monotonic and hence will have \ndc in some
parameter regime. It is worth noticing that in this case the
interaction parameters $g_1$ and $g_2$ both do not scale on the
left wire as it is completely spin polarized while $g_1$ and $g_2$
do scale on the right wire as it is not spin polarized. Hence even
if we begin our \rg flow with symmetric interaction strengths on
both left and right wires, they will develop an asymmetry under
the \rg flow.

Finally, we consider another important aspect that nicely
characterizes these hybrid structures from a spintronics
application point of view. If the \qw on the two sides of the
superconductor are ferromagnetic half metals then we have a
junction of ferromagnet$-$superconductor$-$ferromagnet (\fsfd). We
calculate the tunnelling magnetoresistance ratio
(\tmr)~\cite{review} which is defined as follows
\begin{equation}
\tmr = \left[\dfrac{{G_{\up \up} ~-~ G_{\up \dn}}}{G_{\up
\dn}}\right] \label{tmr}
\end{equation}
Here, $G_{\up \up}$ corresponds to the conductance across the
junction when both left and right wires are in parallel
spin-polarized configurations. $G_{\up \dn}$ corresponds to the case
when the left and right wires are in anti-parallel spin-polarized
configurations. Thus, \tmr is the maximum relative change in
resistance in going from the parallel to the anti-parallel
configuration. For the parallel case, the \car amplitude ($t_A$) is
zero and the only process which contributes to the conductance is
the direct tunnelling process. This is because the \car process
involves non-local pairing of $\up\e$ in the left wire with $\dn\e$
in the right wire to form a Cooper pair. However for $\dn\e$, the
density of states is zero in the right wire which makes this process
completely forbidden. Hence, $G_{\up \up} \propto |t|^2$. On the
other hand, for the anti-parallel case, $G_{\up \dn} \propto
-|t_A|^2$ as there is no density of states for the $\up\e$ in the
right lead and so no direct tunnelling of $\up\e$ across the
junction is allowed; hence \car is the only allowed process. Note
that the negative sign in $G_{\up\dn}$ leads to a very large
enhancement of \tmr (as opposed to the case of standard
ferromagnet$-$normal metal$-$ferromagnet (\fnfd) junction) since the
two contributions will add up.  {A related set-up has been studied
in \cite {giazotto} where also a large \tmr has been obtained.}

One can then do the \rg analysis for both parallel and
anti-parallel cases. It turns out that the equations for $|t|$ and
$|t_A|$ are identical leading to identical temperature (bias)
dependance. The \rg equation for $|t_A|$ is
\bea \frac{dt_A}{dl} = - \beta\, t_A \, \left[1-|t_A|^2\right]
\eea
\noindent Here, $\beta=(g_2 - g_1)/ 2 \pi \hbar v_F$. $|t|$
satisfies the same equation. So, in a situation where the reflection
amplitudes at the junction for the two cases are taken to be equal
then it follows from Eq.~\ref{tmr} that the \tmr will be pinned to
its maximum value \ie~magnitude of $\tmr=2$ and the temperature
dependence will be flat even in the presence of inter-electron
interactions.

\section{Conclusions} In this letter, we have
studied both spin and charge transport in \nsnd, \fsnd, and \fsf
structures in the context of \od \qwd. We calculated the
corrections to spin and charge transport arising from
inter-electron interactions in the \qwd. We demonstrated the
possibility for production of pure \scurrent in such hybrid
junctions and analysed its stability against temperature and
voltage variations. Finally, we also showed that the presence of
the \car process heavily enhances the \tmr in such geometries.

\vspace{.2cm}

We acknowledge use of the Beowulf cluster at H.R.I. S.D. was
supported by the Feinberg Fellowship Programme at WIS, Israel.

\bibliographystyle{eplbib} 
\bibliography{myreferences} 
%

\end{document}